\documentclass[twocolumn,showpacs,superscriptaddress]{revtex4}

\usepackage{amsmath}
\usepackage{amssymb}
\usepackage{graphicx}

\newcommand{\hvcra}{HVC${}_\textrm{RA}$}
\newcommand{\cai}{\ensuremath{\left[{\rm Ca}\right]_i}}

\begin{document}

\title{Stable Propagation of a Burst Through a
 One-Dimensional Homogeneous Excitatory Chain
 Model of Songbird Nucleus HVC}

\author{MengRu Li}
\affiliation{Department of Physics, Duke University, Durham
NC 27708-0305}

\author{Henry Greenside}
\affiliation{Department of Physics, Duke University, Durham
NC 27708-0305}

\date{\today}

\begin{abstract}
  We demonstrate numerically that a brief burst
  consisting of two to six spikes can propagate in a
  stable manner through a one-dimensional homogeneous
  feedforward chain of non-bursting neurons with
  excitatory synaptic connections. Our results are
  obtained for two kinds of neuronal models, leaky
  integrate-and-fire (LIF) neurons and Hodgkin-Huxley
  (HH) neurons with five conductances.  Over a range of
  parameters such as the maximum synaptic conductance,
  both kinds of chains are found to have multiple
  attractors of propagating bursts, with each attractor
  being distinguished by the number of spikes and total
  duration of the propagating burst.  These results
  make plausible the hypothesis that sparse
  precisely-timed sequential bursts observed in
  projection neurons of nucleus~HVC of a singing zebra
  finch are intrinsic and causally related.
\end{abstract}

\maketitle

\section{Introduction}
\label{sec:intro}

A question of great interest to neurobiology is how
animals learn to generate temporal patterns of muscle
activation. An example that has been much studied in
recent years because of its relevance to human
speech~\cite{Doupe99} and because of the rich variety
of possible experiments is how songbirds learn to sing
a song by auditory-guided vocal
feedback~\cite{Catchpole95}. A young male bird first
memorizes the song of an adult male of the same
species. Then over many months, over many iterations
(more than 50,000 iterations for a zebra
finch~\cite{Johnson02}), and by just listening to his
own vocalizations, a young male learns to vary the
activation of its respiratory and syringeal muscles
until his song is able to match accurately the original
memorized song. This is an impressive feat given that
an adult song might last several seconds, that there is
auditory structure that lasts less than 10~ms, that
many syringeal and respiratory muscles need to be
coordinated, and that some species of songbirds are
able to learn and sing many different songs. Nearly all
details of this process remain poorly understood, in
particular how the young male memorizes a complex song,
and how auditory feedback is used to adjust the pattern
of muscle activation until the songbird is able to
reproduce accurately the memorized song.

\begin{figure}
  \centering
  \includegraphics[width=3in]{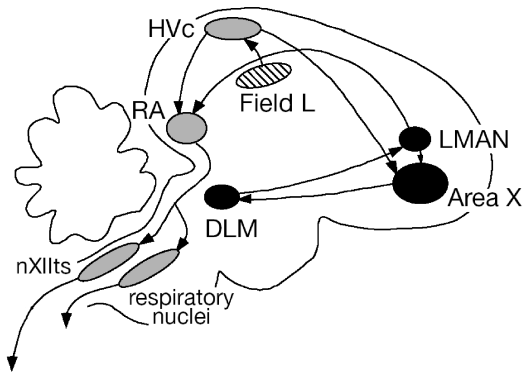}
  \caption{ Schematic anatomical diagram of a sagittal
    cross section of one hemisphere of a zebra finch
    brain, showing the major nuclei associated with the
    learning and production of song.  The nucleus~HVC
    plays a central role since it receives auditory and
    other input, it drives the premotor
    nucleus~RA~\cite{Hahnloser02}, and it sends
    information to a parallel anterior forebrain
    pathway (nuclei~X, DLM, and~LMAN) that play a role
    in the learning and maintenance of a song. There
    are two HVC~nuclei, one on the left and one on the
    right side of the songbird's brain.}
  \label{fig:songbird-brain}
\end{figure}

Studies of songbirds have shown that certain
anatomically and physiologically distinct brain regions
called nuclei are associated with the recognition,
learning, and production of song (see
Fig.~\ref{fig:songbird-brain}). A recent experiment by
Hahnloser et al~\cite{Hahnloser02} recorded
extracellular action potentials (spikes) from neurons
in awake singing male zebra finches and found that the
neurons in the nucleus~HVC\cite{Reiner04} that project
to the robust nucleus of the arcopallium (abbreviated
as~RA~\cite{Reiner04}) have the remarkable properties
of firing sparsely and precisely during singing. (In
the following, HVC~neurons that project to~RA will be
abbreviated as \hvcra~neurons.)  Typically, each
\hvcra~neuron fires a brief burst of about 3-4 spikes
once per song motif with each burst lasting about 6~ms.
(A motif is a cluster of distinct auditory syllables
that is repeated as a single pattern and that lasts an
average of 0.6~s for zebra finch songs.)  Measurements
during successive motifs from a given adult male bird
show that each burst from a particular \hvcra~neuron is
aligned with certain acoustic features of the motif to
a precision of about 0.7~ms~\cite{Hahnloser02}. These
bursts are important to understand since they are
believed to provide the temporal framework for
organizing the syllables of a song~\cite{Fee04}.

The observation that each \hvcra~neuron bursts just
once per motif and other data which show that there is
a reproducible temporal ordering of \hvcra~neurons
according to when they fire their single burst per
motif~\cite{Hahnloser02}, suggest the hypothesis that
bursts propagate along a feedforward network of
excitatory \hvcra~neurons~\cite{Fee04}. The idea is
similar to a path of dominoes such that the falling of
one domino causes the next domino in the path to
topple. Here, a given \hvcra~neuron receives one or
more bursts from other \hvcra~neurons that have fired
recently during the motif. In response, the given
neuron itself fires a burst that helps to activate
neurons that fire later during the motif. That each
\hvcra~neuron fires exactly once per motif implies that
the connections are feedforward (although this
assumption can be weakened if a recurrent burst arrives
while some earlier neuron is still in a refractory
state~\cite{Jin05}). The bursts continue to propagate
until the chain ends and the motif stops, or perhaps
there is a branch point to a different chain whose
dynamics generate a different motif.

While there has been much previous theoretical research
concerning when spikes can propagate along excitatory
networks of different
kinds~\cite{Abeles91,Golomb97,Diesmann99,Reyes03,Litvak03,Osan04,Vogels05},
the songbird experiments on~HVC of Hahnloser et al
during singing~\cite{Hahnloser02} and related
experiments by Mooney of bursts in HVC projection
neurons during audition~\cite{Mooney00} motivate asking
new theoretical questions about the propagation of a
burst through various kinds of networks.

In this paper, we show that a minimal feedforward
architecture---a one-dimensional (1d), homogeneous,
excitatory chain of non-bursting neurons---can support
the stable propagation of a brief high-frequency burst.
By studying two kinds of chains, one with leaky
integrate-and-fire neurons (abbreviated as LIF~neurons)
and one with more realistic single-compartment
conductance-based neurons (abbreviated as HH~neurons
for Hodgkin-Huxley~like), we show that a burst can
propagate in a stable manner that does not require a
careful choice of neuronal model nor a careful tuning
of model parameters.  Depending on the values of
parameters such as the maximum synaptic conductance, we
find that each kind of 1d~chain (LIF or~HH) has
multiple attractors that differ in the number of spikes
and in the total width of the propagating burst.

Our calculations are the first to demonstrate that a
brief high-frequency burst similar to those observed
in~HVC can propagate in a stable manner along a simple
excitatory chain of neurons. (A similar result was
announced independently by Jin and
collaborators~\cite{Jin05}, who studied a more
complicated model. We discuss this preprint briefly in
Section~\ref{sec:conclusions} below.) Our results make
plausible one of the simplest explanations of the
Hahnloser et al result, namely that the observed bursts
of \hvcra~neurons during singing are intrinsic to~HVC
in that external input from other brain regions is not
needed to generate the bursts (except the first burst),
and in that the bursts are causally related such that
one burst initiates the generation of the next
burst. Our results also make some predictions such as
the existence of multistability (different kinds of
propagating bursts can occur in the same chain
depending on how the chain is activated), and that
transitions from one kind of propagating burst to
another kind (differing in the number of spikes and
burst width) can occur as parameters are varied. New
experimental studies of~HVC, especially using optical
methods with high time resolution~\cite{Ohki05} that
can examine the spatiotemporal activity of many neurons
simultaneously, may help to confirm these ideas.

The rest of this paper is organized as follows.
Section~\ref{sec:methods} discusses some details of the
two classes of mathematical neurons that we use in the
1d~chains, and how the excitatory synaptic currents are
modeled.  Section~\ref{sec:results-discussion} presents
numerical results mainly for a 1d~chain of~HH neurons
(as opposed to LIF~neurons), especially the existence
of different attractors corresponding to different
kinds of propagating bursts.  The paper concludes in
Section~\ref{sec:conclusions}, where key results are
summarized. We also discuss there how our results are
related to alternative hypotheses such as that the
bursting arises from propagation through synfire
chains~\cite{Abeles91} or by a central pattern
generator (abbreviated as CPG).

In a second paper~\cite{Li06II,Li06thesis}, we will
discuss how noise and network heterogeneities affect
the propagation of bursts through a synfire chain. This
second paper shows that our key conclusions still hold:
a brief high-frequency burst can still propagate in a
stable manner for a range of noise strengths and for
various amounts of heterogeneity. However, we also find
that single spikes, even if synchronized across a
synfire pool, do not propagate in a stable way for most
of the parameters studied, which suggests that short
bursts of several spikes are important for achieving
robust propagation through realistic synfire chains.

\section{Methods}
\label{sec:methods}

In this section, we discuss some experimental data that
justifies the study of a one-dimensional chain of
excitatory neurons, and we also point out some of the
experimental details that are ignored in our model. We
also discuss some details of the integrate-and-fire and
conductance-based neuronal models with excitatory
synapses that we use in the
1d~chains. Appendix~\ref{appendix:equations-parameters}
provides further details of equations and parameter
values.

\subsection{A Homogeneous 1d~Chain of Excitatory Neurons}

\label{sec:1d-chain-motivation}

The physiological properties of HVC~neurons and how
they are interconnected within~HVC are poorly
understood~\cite{Kubota98,Dutar98,Mooney00,Mooney05,Wild05}
so that it is not possible at this time to develop a
quantitative model of the HVC~microcircuitry, say at
the level of hippocampus
models~\cite{Traub99}. Researchers have
shown~\cite{Dutar98,Mooney00} that there are at least
three main classes of neurons: excitatory neurons that
project to nucleus~RA, excitatory neurons that project
to area~X in the anterior forebrain pathway, and
inhibitory interneurons~\cite{Dutar98,Mooney00} that
connect only to other neurons within~HVC. Recent
paired-electrode recordings by Mooney and
Prather~\cite{Mooney05} have shown that each type
of~HVC neuron makes local connections with the other
two types of HVC~neurons but details of the connections
such as the number, kinds, and strengths of synapses
are incompletely known.

Of special importance for this paper is the
experimental observation that \hvcra~neurons synapse
with other \hvcra~neurons~\cite{Mooney05}. Thus it is
possible for a feedforward network of excitatory
neurons to exist in~HVC, although we emphasize that
there is no evidence presently for such a
network. Further, there are of order~40,000
\hvcra~neurons in HVC~\cite{Fee04}, which should be
sufficiently many to create a chain whose dynamics
spans a motif or even several motifs. (Since the
observed bursts in \hvcra~neurons last
6~ms~\cite{Hahnloser02}, a chain of about one hundred
pools of neurons would suffice to span a motif
of~0.6~s, provided that the bursts do not overlap in
time.)

To determine in principle whether the propagation of a
burst can explain the experimental
data~\cite{Hahnloser02,Mooney00}, we study one of the
simplest possible feedforward networks, namely a
one-dimensional homogeneous feedforward chain of
identical neurons such that each neuron connects via a
single identical excitatory synapse with the next
neuron of the chain. The assumption of homogeneity is
not realistic biologically but reduces the parameter
space of the calculations and increases our ability to
understand how the existence and properties of a
propagating burst depend on parameters.

Our 1d~model leaves out many experimental details (we
discuss this further in
Section~\ref{sec:results-discussion} below), of which
two are especially important, the need for multiple
pathways and the presence of inhibitory neurons.
First, as pointed out by Abeles~\cite{Abeles91} and
others, transmission of information along a
one-dimensional chain is not robust since damage to any
neuron in the chain can alter or stop the transmission
of information. (That~HVC is robust is demonstrated by
the fact that the adult song of a zebra finch changes
little over the life of the bird, even though there is
ongoing neurogenesis and hence a steady turnover of
\hvcra~neurons~\cite{NottebohmJN02,Wang02}.) If some
kind of propagation occurs in~HVC, there must be
parallel pathways.

\begin{figure}[htb]
  \centering
  \includegraphics[height=2.5in]{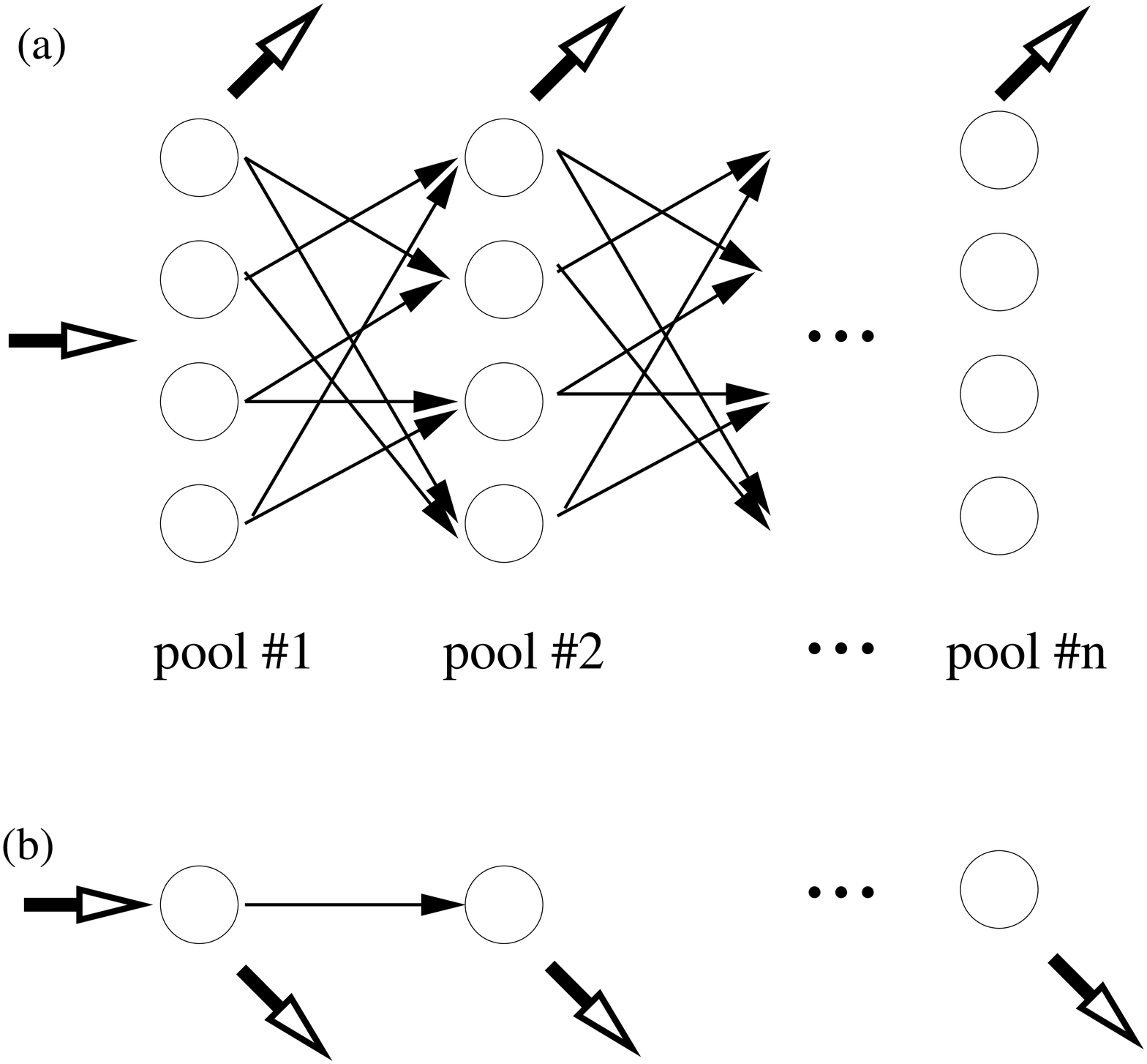}
  \caption{ (a) Schematic diagram of a feedforward
  synfire chain consisting of successive pools of
  neurons (vertical column of circles). Each pool
  contains excitatory \hvcra~neurons that fire
  approximately synchronously to activate neurons in
  the next pool. The horizontal hollow arrow on the
  left represents input that can initiate the synfire
  chain. The hollow arrows at the top of each pool
  indicate efferents that could convey information from
  a given pool to other brain areas. (b) Schematic
  diagram of the one-dimensional homogeneous
  feedforward excitatory chain that we study in this
  paper. In the biologically unrealistic but
  theoretically convenient case that all neurons in the
  synfire chain of~(a) are identical, that all neurons
  in one pool connect to neurons in the next pool in
  the same way with identical synapses, that axonal
  delays between pools can be ignored, and there is no
  noise, the 1d-chain will have dynamics identical to a
  synfire chain whose spikes have become fully
  synchronized. }
  \label{fig:syn-fire-chain}
\end{figure}

A widely studied example of a feedforward neuronal
architecture with multiple pathways is a synfire
chain~\cite{Abeles91,Diesmann99,Ikegaya04}, see
Fig.~\ref{fig:syn-fire-chain}(a). Although there are
variations of this architecture, a synfire chain is
usually described as a strictly feedforward chain of
pools of neurons such that the neurons in one pool
project only to neurons of the next pool. Reliable
transmission occurs even in the presence of noise or of
network heterogeneities because of the multiple paths,
and also because each neuron receives spikes from many
neurons of the previous pool and the spikes
spontaneously become highly synchronized as propagation
continues~\cite{Abeles91,Diesmann99}. The arrival of
synchronized spikes at the many synapses of a given
synfire neuron causes a large postsynaptic current that
triggers new spikes with a probability that can
approach certainty.

The idea that the sparse brief bursts observed in
\hvcra~neurons are causally connected by sequential
firing through some kind of chain has been proposed by
experimentalists~\cite{Fee04,Leonardo05} and assumed in
several theoretical
studies~\cite{Doya98,Abarbanel04jn,Berger04}. In a
recent experiment~\cite{Ikegaya04}, calcium imaging and
timing studies of intracellular recordings have
suggested the existence of synfire chains in the visual
cortices of mice and cats but no similar imaging study
has yet been carried out for songbirds.

The dynamics of our one-dimensional homogeneous chain
is equivalent to the dynamics of a homogeneous
noiseless synfire chain (homogeneity here means equal
numbers of identical neurons per pool all connected in
identical ways from one pool to the next) such that all
spikes within a given pool have become perfectly
synchronized. This is an important observation since
the synapse that connects one neuron to the next in the
1d~chain of Fig.~\ref{fig:syn-fire-chain}(b) then has
to be strong to correspond to the many synapses of a
neuron in the full synfire chain. In general, noise,
heterogeneities, delay times, and imperfect
synchronization of spikes cause the dynamics of a
synfire chain to differ from that of our 1d~chain so
that simulations of real synfire chains are needed to
understand these effects on the propagation of a
burst~\cite{Jin05,Li06II,Li06thesis}.

A second important experimental detail that we leave
out of our model is the presence of the inhibitory
HVC~interneurons. Our justification for this assumption
is simply that the function of these interneurons is
not known. Proposed mathematical models for sequence
generation and sequence recognition incorporate
inhibitory neurons in various
ways~\cite{Kleinfeld89,Drew03,Huerta04,Abarbanel04jn}
but there is no support yet for these models in the
context of the songbird system. We argue that ignoring
the inhibitory neurons is a useful first step toward
understanding the HVC~microcircuitry since the
calculations discussed below show that a minimal
idealized 1d~chain of identical excitatory neurons
already suffices to produce a stable propagating burst,
in which case other effects such as noise,
inhomogeneities, and interneurons could be included
perturbatively. Further, some experiments suggest that
ignoring the interneurons may not be too severe an
assumption.  HVC~interneurons tend to fire rapidly,
chronically, and in approximate synchrony throughout a
motif~\cite{Fee04}, and rapidly and chronically during
audition of a bird's own song~\cite{Mooney00}, so that
the interneuron spikes might not be closely correlated
with the brief sparse bursts of the \hvcra~neurons. In
this case, the main purpose of the interneurons might
to prevent runaway excitation of the projection
neurons.

We conclude this section with the technical observation
that our assumption that the 1d~chain is homogeneous
and that there is no delay along axons (the axons along
the chain have effectively zero length) allow the
1d~chain to be simulated by integrating numerically a
\textit{single} neuron with a single afferent
synapse. During the integration, the output of the
neuron can be stored in memory and then used as input
to the same neuron which then represents the next
neuron in a chain. A related idea was used by
Reyes~\cite{Reyes03}, who used a single real
hippocampal neuron in a slice together with the dynamic
clamp technique to simulate a synfire chain consisting
of many pools of identical neurons.

\subsection{A Leaky Integrate-and-Fire (LIF) Model of an \hvcra~Neuron}

We studied propagation of bursts along 1d~chains
consisting of one of two types of neurons, leaky
integrate-and-fire (LIF) neurons that are described in
this section, and more realistic conductance-based
Hodgkin-Huxley (HH) neurons that are described in the
next section. Neither neuronal model bursts
intrinsically when subject to a direct current (DC)
external stimulus. Provided that an initial burst is
applied to the first neuron of the 1d~chain, we find
that successive neurons are capable of
generating an identical output burst when stimulated
itself by a burst.

With respect to conductance-based models, LIF
models~\cite{Dayan01} are attractive since they have
fewer parameters, lead to efficient numerical
simulations, and are sometimes amenable to analytical
methods. The $k$th~LIF neuron of the 1d~chain ($k \ge
1$) was modeled as a first-order non-autonomous
ordinary differential equation
\begin{equation}
\label{eq:lif}
  \tau_m \frac{dv^k}{dt} = v_0 - v^k  
   +  R \left( I^k_e(t) +  I^k_S(t) \right) ,
\end{equation}
where the parameter~$\tau_m$ is the capacitive
$RC$-time scale of the neuronal membrane, the
variable~$v^k(t)$ is the potential difference across
the membrane of the $k$th~neuron in the chain, the
external current~$I^k_e(t)$ of the $k$th~neuron is a
specified function (in this paper applied only to the
first neuron to initiate activity in the chain), the
synaptic current~$I^k_S(t)$ arises from spikes in the
previous $(k-1)$st~neuron of the chain (see
Eqs.~(\ref{eq:total-synaptic-current})
and~(\ref{eq:lif-syn-current}) below), the
parameter~$v_0$ is the resting potential toward which
the potential~$v$ asymptotes in the absence of external
and synaptic currents ($I_e=I_S=0$), and the
parameter~$R$ is the total resistance of the membrane.

Eq.~(\ref{eq:lif}) was supplemented with the usual
spiking rule that, whenever the potential~$v^k(t)$ is
increasing and crosses a specified threshold
value~$v_\textrm{thresh} > v_0$, the neuron is assumed
to have spiked and the potential~$v^k$ is instantly and
discontinuously decreased to a reset value
$v_\textrm{reset} < v_0$. An absolute refractory period
was included in this LIF~model by freezing the
potential~$v^k$ to the value~$v_\textrm{reset}$ for a
time interval $t_\textrm{refract}$ after a spike.  We
found that including a refractory interval of
$t_\textrm{refract}= 1\,\rm ms$ produced only minor
qualitative changes to the results obtained in the
absence of a refractory period, for example a 1~ms
refractory period expanded by a modest amount the
basins of attraction for the different stable
propagating bursts discussed in
Section~\ref{sec:results-discussion} (see
Fig.~\ref{fig:steadyPlot} below). Although
\hvcra~neurons show some accommodation~\cite{Mooney00},
we did not include this detail in our LIF~model.

The total postsynaptic current~$I^k_S(t)$ of the
$k$th~neuron in the chain was assumed to be a linear
accumulation
\begin{equation}
   I^k_S(t) = \sum_i I_s(t-t^{k-1}_i) ,
   \label{eq:total-synaptic-current}
\end{equation}
of postsynaptic currents~$I_s(t-t^{k-1}_i)$
associated with spikes that occurred in the previous
${(k-1)}$st~neuron of the chain at times~$t^{k-1}_i <
t$. (Eq.~(\ref{eq:total-synaptic-current}) also assumes
that there are no time delays arising from the time for
a spike to propagate from one neuron to the next.) The
time dependence of the synaptic conductivity associated
with a single spike at time~$t^k_i=0$ was modeled as a
double exponential~\cite{Dayan01}, leading to the
following expression for the postsynaptic current per
spike:
\begin{equation}
   I_s(t,v) = n I_0 \left( 
    e^{\displaystyle -t/\tau_1} -  e^{\displaystyle
    -t/\tau_2}
    \right) .
   \label{eq:lif-syn-current}
\end{equation}
The slow and fast time constants~$\tau_1$ and~$\tau_2$
had respectively the values~1.1~ms and~0.2~ms to match
roughly the~4.0~ms onset-to-peak time constants
observed in paired intracellular recordings of
\hvcra~neurons~\cite{Mooney05}.  The
value~$I_0=0.3\,\rm nA$ was chosen to give a 1.0~mV
peak excitatory post-synaptic potential (EPSP) which is
consistent with experiment~\cite{Mooney05}.  The
number~$n$, which varied from~1 to~32, indicated the
synaptic strength in terms of the number of synchronous
spikes that a neuron would receive if placed in a
uniform synfire chain with~$n$ neurons per pool and all
neurons of one pool connecting to all neurons in the
next pool (with synapses of the same form
Eq.~(\ref{eq:lif-syn-current})). The range~1 to~32 was
found empirically to span the range of stable dynamics,
see Fig.~\ref{fig:steadyPlot} below.

An homogeneous chain was then obtained by using for
each LIF~neuron the same ten parameters~$\tau_m$,
$v_0$, $R$, $v_\textrm{thresh}$, $v_\textrm{reset}$,
$t_{\rm refract}$, $n$, $I_0$, $\tau_1$, and $\tau_2$,
and the same function Eq.~(\ref{eq:lif-syn-current})
for the postsynaptic current per spike. A time
constant~$\tau_m = 15\,\rm ms$ was used to approximate
the {\it in vivo} response time~\cite{Mooney00}, and a
membrane resistance of~$R=60\,\rm M\Omega$ was chosen
to match the impedance of
\hvcra~neurons~\cite{Mooney00}.  Other parameter values
used were~$v_0=-70$~mV, $v_\textrm{thresh}=-55$~mV,
$v_\textrm{reset}=-75$~mV, and~ $t_{\rm
refract}=1.0$~ms.

The chain of LIF~neurons represented by
Eqs.~(\ref{eq:lif})-(\ref{eq:lif-syn-current}) was
integrated by using a forward-Euler
method~\cite{Kincaid96} with a constant time step
of~$\Delta{t}=0.01\,\rm ms$.  (Smaller time steps by
factors of four did not lead to significant changes in
the results.) The supplementary reset and refractory
rules were applied at the end of each time step. For
most runs, a zero external current~$I^k_e=0$ was
assumed for each neuron of the chain except for the
first neuron, for which a step function was used to
stimulate a burst.  The code was programmed and run
using the computational mathematics program Matlab
version~6.5~\cite{Matlab}.

\subsection{A  Single-Compartment Hodgkin-Huxley
  Model of an \hvcra~Neuron}

\label{sec:hh-model-of-hvcra-neuron}

The second neuronal model that we used to study the
propagation of a burst in a 1d~chain of excitatory
neurons was a single-compartment model based on the
Hodgkin-Huxley equations with five representative
conductances.  The evolution equation for the membrane
potential~$v^k(t)$ of the~$k$th neuron in the chain was
\begin{equation}
  C_m {dv^k \over dt} = \sum_{i=1}^5 g_i(t,v^k)
   \left( v_i - v^k \right)  +  I^k_e(t) + I^k_S(t) .
  \label{eq:HH-model}
\end{equation}
The symbols have the following meanings: $C_m$ is the
total membrane capacitance, $g_i(t,v)$~is the
voltage-dependent conductance of the $i$th~kind of
membrane channel, $v_i$~is the resting potential for
the $i$th~channel, and the currents~$I^k_e(t)$
and~$I^k_S(t)$ have the same meaning as in
Eq.~(\ref{eq:lif}). Appendix~\ref{appendix:equations-parameters}
gives the details of parameter values and other
evolution equations related to the conductances~$g_i$.

Although more realistic in terms of its time
dependence, the HH~model Eq.~(\ref{eq:HH-model}) is not
necessarily more scientifically appropriate than
LIF~models since the properties of the various membrane
conductances are only partially known for
HVC~neurons~\cite{Kubota91,Kubota98,Dutar98,Mooney00},
the spatial distribution of the channels in HVC~neurons
has not been determined (which would be needed to
construct accurate multi-compartment models), and
little is known about the type, strengths, and
locations of synapses in~HVC. We did find that a
single-compartment HH~neuron with five conductances is
only able to match some features of the
experimental~HVC data, and it is not clear which
experimental details are more important than others to
include in the process of fitting a mathematical model
to data. For these reasons, and because this paper is
concerned whether in principle an HVC-like burst can
propagate stably through an excitatory chain, the
conductances of the HH~model Eq.~(\ref{eq:HH-model})
were only loosely based on existing HVC~data.

Our starting point for choosing the conductances in
Eq.~(\ref{eq:HH-model}) was a recent paper by Prinz et
al~\cite{Prinz03}, which showed that an
eight-dimensional phase space obtained by varying the
maximum conductances of eight channels (whose
functional properties were obtained from lobster
stomatogastric neuronal data) contained a great
diversity of dynamical behavior. We chose the
functional forms of the leakage, sodium, and potassium
channels from this paper and added two other channels
as suggested by current-clamp data of~HVC
neurons~\cite{Kubota91}: a fourth channel was a
low-threshold transient calcium current, and a fifth
channel was a calcium-activated potassium current
(denoted in the following by the symbol~KCa).  The
calcium channel was chosen to activate around -50~mV
and inactivate around -80~mV with time scales such that
the channel could be activated
transiently~\cite{Kubota91,Huguenard92}.
The KCa~conductance activates the potassium current
after intracellular calcium concentration rises, with a
functional form adopted from Yamada's
model~\cite{Yamada98}.  The maximum conductances of all
five conductances were adjusted iteratively by hand to
produce a short spike width and to generate a spike
adaptation with a high initial firing rate as observed
by Kubota~\cite{Kubota91}.
A reasonably good fit to the Mooney {\it in~vivo} data
(see Fig.~1 of Ref.~\cite{Mooney00}) unfortunately
could not also match Kubota's {\it in~vitro} spike
profiles of Ref~\cite{Kubota98} and vice versa. (We
hope to obtain more complete fits to HVC~data by adding
more channels and more compartments when more
anatomical and physiological data become available.)


The total synaptic current was also assumed to satisfy
the linear relation
Eq.~(\ref{eq:total-synaptic-current}) but an
alpha~function~\cite{Dayan01} was used instead of
Eq.~(\ref{eq:lif-syn-current}) to approximate the
time-dependent probability of a channel opening.  The
excitatory postsynaptic current (EPSC) arising from a
spike that occurs at time~$t=0$ had the form
\begin{equation}
   I_s(t;v) = 
    g_s C \left( t \over \tau_s \right) e^{\displaystyle -t/\tau_s} 
   \left( v_s - v \right) ,
  \label{eq:hh-syn-current}
\end{equation}
where the constant~$C$ with value~$e$ normalizes the
maximum value of the expression
$C(t/\tau_s)\exp(-t/\tau_s)$ to~1. This normalization
makes the parameter~$g_s$ the maximum conductance,
which we varied over the range 0 to 0.55~nS.  In
Eq.~(\ref{eq:hh-syn-current}), the synaptic time
constant~$\tau_s$ was fixed with value~7~ms and
the synaptic reversal voltage~$v_s$ was set to~120~mV
to make the synapse excitatory.
Eqs.~(\ref{eq:HH-model}),
(\ref{eq:total-synaptic-current})
and~(\ref{eq:hh-syn-current}) together with the
evolution equations for gate variables given in
Appendix~\ref{appendix:equations-parameters} were
integrated with the Neuron simulation code
version~5.6~\cite{NeuronCode}, with a constant time step
of~$\Delta{t}=0.1\,\rm ms$.  




\section{Results and Discussion}

\label{sec:results-discussion}

\subsection{Results for a 1d Homogeneous Chain of HH~Neurons}

The calculations discussed below of one-dimensional
homogeneous chains using the LIF or HH~neurons of the
previous section with excitatory synapses show that a
brief high-frequency burst similar to that observed in
\hvcra~neurons during singing~\cite{Hahnloser02} can
propagate stably over a range of parameter values.
These results make more plausible the hypothesis that
an excitatory chain of \hvcra~neurons generates the
sparse precisely-aligned bursts observed
experimentally~\cite{Hahnloser02}. Our results show
further that different asymptotic attractors can exist
for given parameter values so that hysteresis can occur
(different initial conditions can lead to different
non-transient dynamics).  The fact that the bursts
exist as an attractor means that there is a transient
time during which properties of the burst such as the
number of spikes and burst width (alternatively, the
average burst frequency) evolve until the final stable
values are obtained. This transient time varies with
the initial state used to start the chain and with the
choice of parameters.

Since these qualitative conclusions turned out to be
similar for LIF and HH~neuronal models, we report here
results mainly for the HH~models of
Section~\ref{sec:hh-model-of-hvcra-neuron} and mention
briefly in the next section how the results differ for
a chain of~LIF neurons. Since the parameter spaces for
LIF and HH~neurons are high-dimensional (10~parameters
for~LIF, about 25~parameters for the HH~neurons, and
these do not include the choice of the functional
form~$I_s(t,v)$ for the~EPSC per spike), to establish
our key results we varied only one parameter
systematically, namely the maximum synaptic
conductance~$g_s$ of Eq.~(\ref{eq:hh-syn-current}) for
HH~neurons, or equivalently the number of synchronous
presynaptic spikes~$n$ of
Eq.~(\ref{eq:lif-syn-current}) for LIF neurons.

\begin{figure}[tb]
  \includegraphics[height=2.7in]{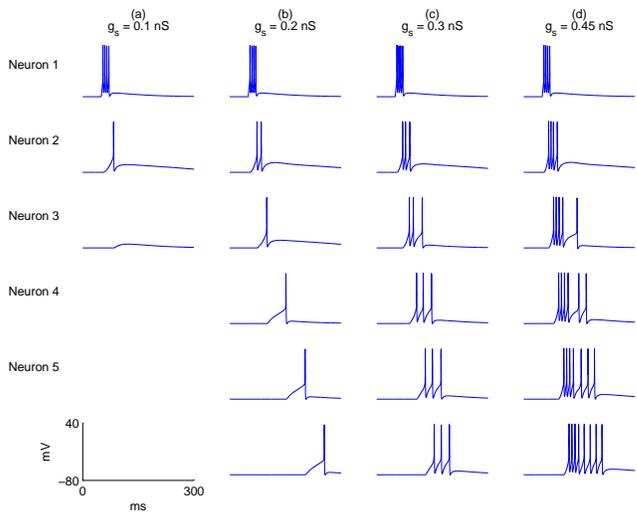}
  \caption{ Propagation of an initial 5-spike burst
    through an homogeneous one-dimensional excitatory
    chain of HH~neurons, for different maximum synaptic
    conductances~$g_s$ of
    Eq.~(\ref{eq:hh-syn-current}). Each column is
    corresponds to a fixed~$g_s$ value, and successive
    rows of a given column show the membrane
    potential~$v^k(t)$ as a function of time for
    successive neurons in the chain ($k=1, 2,
    \cdots$). {\bf (a)}~For a weak synaptic coupling
    strength~$g_s=0.1\,\rm nS$, the initial burst
    decays by the third neuron.  {\bf (b)}~For a
    stronger coupling~$g_s=0.2\,\rm nS$, the initial
    bursts decays to an invariant single spike.  {\bf
    (c)}~For~$g_s=0.3\,\rm nS$, the initial burst
    evolves to a stable state with three unevenly
    spaced spikes.  {\bf (d)}~For still stronger
    couplings $g_s \ge 0.45\,\rm nS$, the initial burst
    can be unstable for some initial states and the
    number of spikes increases steadily.}
  \label{fig:burst-propagation}
\end{figure}

Figure~\ref{fig:burst-propagation} shows how the same
initial burst of five spikes propagates through a
1d~excitatory chain of HH~neurons for several different
values of the maximum synaptic conductance~$g_s$. The
initial burst was created by using an external
current~$I^1_e(t)$ of Eq.~(\ref{eq:HH-model}) for the
first neuron in the chain to inject a square pulse of
current that caused five spikes to appear in rapid
succession.  For a synaptic coupling strength~$g_s$
smaller than about~$0.1\,\rm nS$,
Fig.~\ref{fig:burst-propagation}a shows that the
initial burst rapidly dies out by the third neuron of
the chain and all spikes disappear. Over a range of
stronger couplings $0.2 < g_s < 0.4\,\rm nS$, the
initial bursts evolves into a stable invariant
propagating burst that can have one to five spikes
(columns (b) and~(c) of
Fig.~\ref{fig:burst-propagation}). For stronger
couplings $g_s > 0.4\,\rm nS$, the initial burst is
unstable and the number of spikes grows steadily
without limit. However, other initial conditions can
lead to stable bursts in this range, an example of
hysteresis.

\begin{figure}[tb]
  \centering
  \includegraphics[height=2.7in]{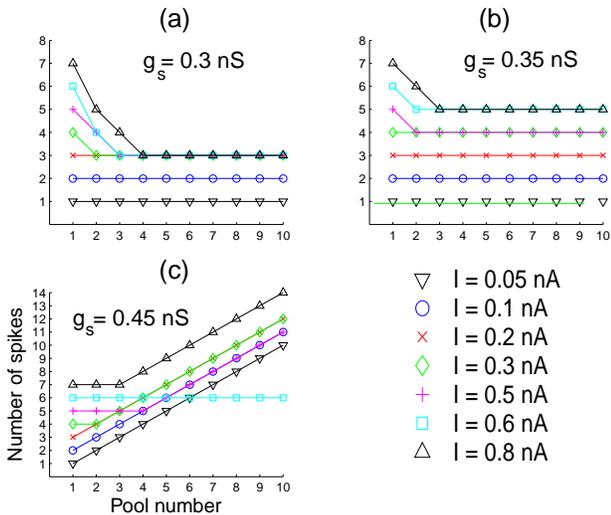}
  \caption{(Color onlin) The dependence of transient dynamics and
    resulting attractors of propagating bursts on the
    initial injecting current magnitude~$I^1_e(t)$ and
    on the synaptic conductance~$g_s$. For weak
    couplings with $g_s < 0.2\,\rm nS$, all initial
    states decay to zero (not shown here). {\bf (a)}
    For~$g_s=0.3\,\rm nS$, there are multiple basins of
    attraction corresponding to asymptotic propagating
    bursts with 1, 2, or~3 spikes. Transients decay
    rapidly by the third neuron of the chain. {\bf (b)}
    Increasing the synaptic strength to~$g_s =0.35\,\rm
    nS$ increases the number of attractors, with final
    bursts containing 1 to~5~spikes.  {\bf (c)} For a
    larger synaptic conductance~$g_s=0.45\,\rm nS$,
    initial conditions lead to stable or unstable
    states. An initial burst with 6~spikes will evolve
    to a slightly different invariant burst with also
    6~spikes (squares). }
  \label{fig:initCondProp}
\end{figure}

Fig.~\ref{fig:initCondProp} provides a more global
understanding of the transient dynamics and resulting
attractors. In each panel, the vertical axis indicates
the number of spikes observed in a burst while the
horizontal axis indicates the position~$k$ of a neuron
along the chain. For a weak synaptic coupling~$g_s<
0.2\,\rm nS$, all initial states decay to zero spikes
(not shown in the figure).  For a synaptic coupling
of~$g_s=0.3\,\rm nS$ (Fig.~\ref{fig:initCondProp}a),
initial bursts with three or more spikes decay within
four neurons to a common final burst of three
spikes. An initial burst with two spikes evolves
slightly to an invariant burst with also two spikes,
and a similar conclusion holds for an initial burst
with a single spike. As the synaptic strength is
increased to~$g_s=0.35\,\rm nS$
(Fig.~\ref{fig:initCondProp}b), the number of
attractors increases so that stable bursts with one to
five spikes are observed depending on the initial state
of the first neuron. The transient time is still rather
short, with the final number of spikes stabilizing in
all cases within three neurons ($k \le 3$). Finally for
synaptic couplings stronger than about~$0.4\,\rm nS$
(Fig.~\ref{fig:initCondProp}c), stable and unstable
states are observed depending on the initial state of
the first neuron. The unstable bursts all grow at the
same rate, with an extra spike appearing for each next
neuron traversed.

\begin{figure}[tb]
  \centering
  \includegraphics[height=2.7in]{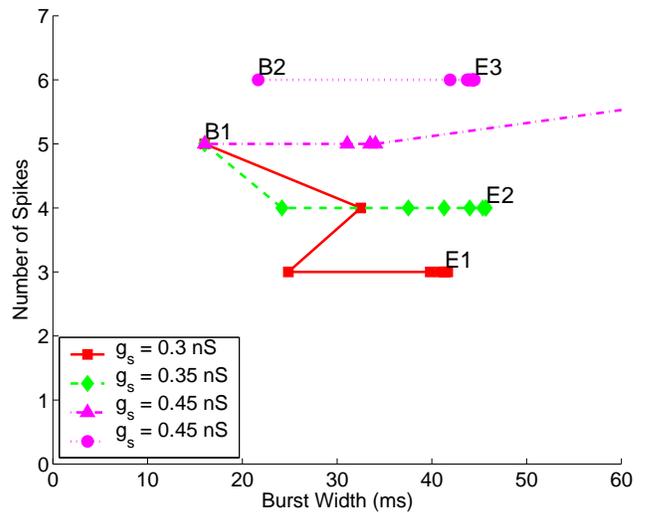}
  \caption{ (Color online) Different synaptic conductances~$g_s$ lead
    to different final propagating burst widths and
    spike numbers, which shows that the total burst
    duration is another important characteristic of the
    final attractor.  For $g_s = 0.3\,\rm nS$, the
    curve of block symbols evolves into the point~E1.
    For $g_s = 0.35 \,\rm nS$, the curve with diamond
    symbols ends in the point~E2.  For $g_s = 0.45\,\rm
    nS$, the propagation is unstable (upper triangle)
    except for the initial condition at point~B2.  }
  \label{fig:transient-states}
\end{figure}

The transient dynamics of Fig.~\ref{fig:initCondProp}
are presented in somewhat finer detail in
Fig.~\ref{fig:transient-states}, which shows that a
non-transient propagating bursts is characterized by at
least two parameters, their total width and the number
of spikes.  For example, the same beginning state~$B_1$
can evolve into three different end points depending on
the synaptic coupling strength~$g_s$: point~$E_1$ with
three spikes, point~$E_2$ with 4~spikes and with a
slightly longer width of~45~ms, and an unstable state
(triangles).

\begin{figure}[tb]
  \centering
  \includegraphics[height=2.7in]{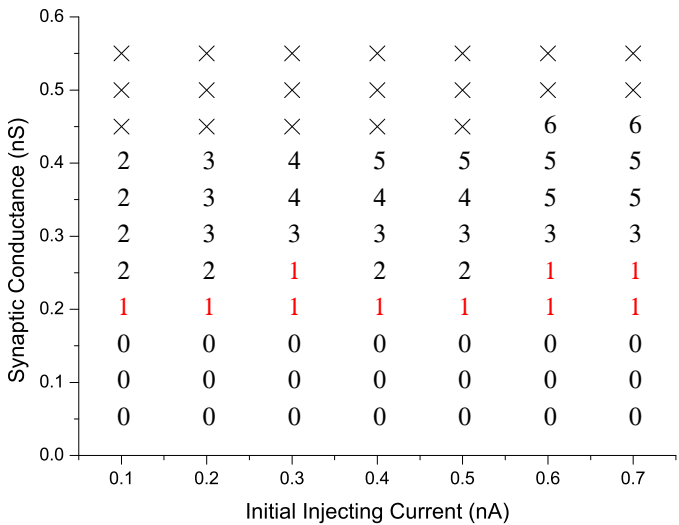}
  \caption{ Plot of the number of spikes observed in a
  final non-transient propagating burst for different
  initial amplitudes of a square current pulse of fixed
  duration~20 ms and for different maximum synaptic
  conductances~$g_s$.  Different basins of attraction
  are observed, with 0-spike and unstable bursts having
  the largest basins, and roughly equal size basins for
  bursts with 1 to~5 spikes in the final burst.}
  \label{fig:steadyPlot}
\end{figure}

Finally, Fig.~\ref{fig:steadyPlot} indicates the
approximate basins of attraction that exist for
different initial current amplitudes (horizontal axis)
and different maximum synaptic conductances (vertical
axis). The plotted numbers indicate the number of
spikes observed in the asymptotic state for the
specified axis values.  This plot shows again how all
states decay away for sufficiently weak synaptic
coupling, all states are unstable for sufficiently
strong coupling, and that stable propagating bursts
with~1 to~6 spikes can be found for intermediate
coupling strengths. The basins of attraction turn out
to be of comparable size except for the tiny basin
corresponding to a 6-spike burst.

\subsection{Results for a 1d Homogeneous Chain of LIF
Neurons}

\label{sec:results-1d-homog-lif-chain}

Our results for homogeneous chains of~LIF neurons were
qualitatively similar to the above results for chains
of HH~neurons and we do not show the corresponding
results.  Instead of initiating the chain with a
current pulse, the first LIF~neuron of the chain was
initiated with a synaptic current corresponding to a
burst of equally-spaced spikes whose times of
occurrences~$t^0_i$ were specified as initial data (see
Eq.~(\ref{eq:total-synaptic-current})).  Similarly,
instead of a maximum conductance~$g_s$ being varied,
the synaptic coupling strength~$n$ of
Eq.~(\ref{eq:lif-syn-current}) was varied between~1
and~32 which we found to span the same kinds of states
discussed in Figs.~\ref{fig:burst-propagation}
and~~\ref{fig:steadyPlot}, from a 0-spike final state
to an initial burst that grew without bound in width
and in the number of spikes.  Because the LIF neurons
have no intracellular calcium dynamics, the number of
spikes in the final state was primarily determined by
the initial spike number. In contrast, the
intracellular calcium current of a HH neuron can boost
depolarization and maintain propagating bursts in
cohesion so that chains of HH~neurons have bigger
basins of attraction for stable bursts.


\subsection{Other Hypotheses and Related Theory}
\label{sec:other-hypothese-and-related-theory}

We discuss briefly here some other hypotheses and
related theoretical work that might explain the data of
Hahnloser et al~\cite{Hahnloser02} and of
Mooney~\cite{Mooney00} but that do not involve the
propagation of bursts through a feedforward network.

While a feedforward network, especially a synfire
chain, is one of the simplest concepts that might
explain sparse precisely-timed bursting, there is much
theoretical work dating back to the~1980s which shows
that a recursive network with inhibitory and excitatory
connections is capable of learning and producing
different kinds of temporal
sequences~\cite{Kleinfeld89,Hertz91}. These networks
generalize the Hopfield attractor model of associative
memory~\cite{Hopfield82,Hertz91} by allowing
non-symmetric couplings between pairs of neurons, and
by using two kinds of synapses, ``fast'' synapses that
stabilize a given network state, and ``slow'' synapses
that cause successive transitions between the
quasistatic network states.  Thus there is no
difficulty in principle for a recursive network to
store, generate, or learn many temporal sequences of
the sort observed in~HVC. (We note that other classes
of recursive models are possible, e.g., Huerta and
Rabinovich~\cite{Huerta04} discuss a model with neurons
that are strictly inhibitory or strictly excitatory,
rather than allow any neuron to have inhibitory and
excitatory connections with other neurons.)
 
Asymmetric Hopfield-like recursive networks have
several attractive features for modeling
HVC~dynamics. The HVC inhibitory
neurons~\cite{Mooney00} can be incorporated in a
natural way since temporal sequences are stored via
neuronal connection strengths~$J_{ij}$ that typically
have positive (excitatory) and negative (inhibitory)
values. Recursive networks that generate sequences can
evolve via a simple Hebbian learning
rule~\cite{Kleinfeld89} from densely interconnected
neurons (see Figs.~2 and~3 of Ref.~\cite{Mooney00}) and
so are possibly easier to form during maturation of
brain tissue than purely feedforward
networks. Asymmetric Hopfield models are naturally
fault-tolerant and error-correcting so that details of
a stored sequence are weakly affected if neurons or
synapses are modified, deleted, or added. Finally, a
single network of this kind is capable of storing and
generating many different temporal sequences, which is
consistent with the ability of some songbirds to learn
and sing many different songs~\cite{Catchpole95}.

However, further work is needed to determine whether
existing recursive models~\cite{Kleinfeld89} are
capable of taking into account specific experimental
details of~HVC such as the fact that an \hvcra~neuron
fires briefly just once during a one-second-long motif
(a neuron is more likely to fire multiple times during
a motif if there are recurrent pathways), that time
interval between successive bursts is short (estimated
to be about 10~ms~\cite{Fee04}, this is possibly too
short for a separation of time scales to exist between
slow and fast synapses~\cite{Kleinfeld89}), that there
is a precise alignment of bursts with auditory features
during singing~\cite{Hahnloser02} and during
audition~\cite{Mooney00}, and especially that
HVC~inhibitory neurons fire tonically while the
HVC~projection neurons fire sparsely during
singing~\cite{Hahnloser02} and during
audition~\cite{Mooney00}. (There is no dynamical
distinction between inhibitory and excitatory synapses
in the recursive models~\cite{Kleinfeld89}.)

Another alternative explanation for the origin and
precise alignment of \hvcra~bursts during singing and
during audition is that these bursts are driven by
precisely timed external inputs. This possibility is
suggested by the fact that other nuclei such as~NIF
and~Uva are known to provide distributed input to HVC,
and to all three classes of
HVC~neurons~\cite{Coleman04,Rosen06}. However, some
experiments suggest that this is a less plausible
explanation than intrinsic generation of bursts
within~HVC. For example, preventing auditory input
to~HVC by destroying a bird's cochlea (which deafens
the bird) or by lesioning nucleus~NIF does not prevent
a bird from singing, nor does it cause a bird's song to
change substantially over time scales of days.
Although extracellular recordings of an awake singing
bird of the kind reported by Hahnloser et
al~\cite{Hahnloser02} have not been carried out after
deafening or lesioning, the fact that the song does not
change immediately suggests that the same pattern of
sparse bursts should still be observed
in~\hvcra~neurons. Since the inputs to~HVC remain
incompletely characterized, external driving, or some
combination of external driving and intrinsic HVC
circuitry, might be able to explain the observed bursts
of \hvcra~neurons during singing.

A variation of the idea of precise external input
driving the bursts in~HVC is the possibility that part
of~HVC acts as a central pattern generator (CPG) that
drives the \hvcra~neurons. For example, the
time-ordering of bursts in \hvcra~neurons observed by
Hahnloser et al~\cite{Hahnloser02} could arise from
a~CPG composed of other HVC~neurons such that the~CPG
connects to the \hvcra~neurons via axons of different
lengths and so with different delay times. (The
dynamics of the~CPG itself could be explained as a
non-symmetric Hopfield model~\cite{Kleinfeld89} or a
cyclic synfire chain.) A CPG-based model would differ
from a synfire chain in that a given burst does not
trigger causally the next burst like dominoes falling
over along some path. For example, destroying all
\hvcra~neurons that burst at a certain time in a
CPG~model would not stop the firing of \hvcra~neurons
that normally would burst a short time later. A recent
\textit{in vitro} slice experiment~\cite{Solis05}
suggests that HVC~has intrinsic oscillatory dynamics
consistent with the existence of a~CPG, but it is not
yet known how these slice data relate to the dynamics
of \hvcra~neurons during song.

\section{Conclusions}
\label{sec:conclusions}

For an idealized one-dimensional homogeneous
feedforward chain of excitatory non-bursting neurons,
we have shown by numerical calculations that a brief
high-frequency burst of two to six spikes can propagate
in a stable way for various choices of parameter
values. Previous studies of excitatory networks have
not addressed the propagation of bursts and so have not
been directly relevant for recent experiments
concerning the properties of nucleus~HVC. Previous
studies~\cite{Diesmann99} have also mainly involved
networks of LIF~neurons (to reduce the computational
effort) and so have not taken into account the more
complex dynamics and realistic time scales of
HH~neurons, especially the effects of a calcium current
which experiments have shown to be present in
HVC~neurons~\cite{Kubota91,Kubota98}. Our results show
that stable bursts exist over a range of parameter
values for both LIF and~HH neuronal models so that the
propagation of stable bursts in a homogeneous chain is
not sensitive to the choice of model or of model
parameters.

These results make plausible the hypothesis that the
sparse precisely-aligned high-frequency brief bursts
observed by Hahnloser et al~\cite{Hahnloser02} in
\hvcra~neurons during singing are intrinsic (do not
need external input) and are causally connected in that
one neuron initiates bursting in the next neuron and so
on until the end of the chain. However, other
explanations of the sparse precise bursts are possible
as discussed in
Section~\ref{sec:other-hypothese-and-related-theory},
and further experiments, especially using methods that
can analyze many neurons at once with good time
resolution, would be valuable in helping to distinguish
feedforward chains from external driving, recursive
networks, or other mechanisms.

We have not addressed in this paper why \hvcra~neurons
burst in the first place in, i.e., why information is
transmitted as bursts rather than as a population of
spikes. (Sparse firing can aid the learning of
sequences as pointed out by Fiete et al~\cite{Fiete04}
but this does not require bursts specifically.) There
is somewhat of a paradox here in that the commonly
posited purpose of a burst is to increase the
likelihood of accurate transmission through a noisy
network~\cite{Lisman97,Prida97}, while a similar goal
is achieved without bursts by using a synfire
chain~\cite{Abeles91}. Thus if a synfire chain is the
correct architecture for \hvcra~neurons, it is less
clear why bursts are needed, especially since HVC~has
much less noise than cortical neurons and HVC~synapses
are substantially stronger than cortical
synapses~\cite{Mooney05}. Unpublished calculations by
the authors~\cite{Li06II,Li06thesis} that extend the
present paper to noisy heterogeneous synfire chains do
suggest that bursts are important for achieving robust
propagation even for a synfire architecture.

We finish this paper by summarizing some consequences
of the chain hypothesis and of our calculations for
recent and future songbird experiments.
\begin{enumerate}
  
\item A simple consequence of the feedforward nature of
  an excitatory chain is that an experimentalist in
  principle should be able to initiate singing of an
  arbitrary contiguous segment of a motif. For example,
  if it is possible to stimulate an intermediate part
  of a chain in such a way that a burst begins to
  propagate toward the end of the chain, a songbird
  might then sing a motif that starts somewhere in the
  middle and that then continues to its end. Similarly,
  physically terminating the chain at some intermediate
  point (say with a local lesion) could cause a motif
  to terminate before its usual endpoint, or some
  combination of stimulating in the middle of the chain
  and lesioning later in the chain could be implemented
  in which case the motif could start somewhere in the
  middle and terminate prematurely.

  Electrical stimulation of~HVC via a single
  extracellular electrode in an awake behaving bird
  does not initiate singing~\cite{Vu94,Vicario95,Vu98}
  but instead resets a motif to its beginning (if the
  stimulus occurs while a bird was singing). The
  failure to initiate the singing of a motif, or part
  of a motif, does not rule out the existence of an
  excitatory chain but instead could imply that a
  carefully-arranged pattern of input spikes might be
  needed to initiate activity at some point in the
  chain, especially if the network is a synfire chain
  for which most neurons in a pool must fire in near
  synchrony for neurons in the next pool to
  fire. Identifying the location of \hvcra~neurons in a
  chain by, say, optical imaging would be a valuable
  prior step that could suggest how to work out an
  electrical or photo-uncaging
  protocol~\cite{Losonczy06} that could initiate a
  chain at an arbitrary point along its length, or
  could suggest how to lesion neurons that would
  terminate propagation of a burst hence a motif
  prematurely.

\item Our calculations suggest that a homogeneous chain
  based on~LIF or HH~neuronal models should generally
  have multiple basins of attraction.  Different kinds
  of asymptotic stable bursts differing in the number
  of spikes and in their total duration should be
  observed in nucleus~HVC for different values of
  synaptic strength and other neuronal parameters.
  Further, hysteresis can be expected such that, for
  identical experimental conditions, different initial
  states will evolve to different kinds of asymptotic
  bursts.
  

  Hysteresis under fixed experimental conditions can be
  explored by initiating different kinds of bursts.
  This may be possible if the neurons at the beginning
  of the chain can be identified and an appropriate
  stimulation protocol worked out (see points~1 and~3).

  Most nonlinear network models, including asymmetric
  Hopfield models and CPGs, also have multiple basins
  of attraction so this is not a distinguishing
  property of chain models. However, the number and
  types of basins of attraction, and how these basins
  vary under the influence of pharmaceuticals that alter
  synaptic strengths, may lead to predictions that
  distinguish feedforward from recursive models.
  
\item Chains of neurons, like the neuronal models that
  make up the chains, are driven dissipative systems so
  that initial states will generally evolve through a
  transient before settling into an asymptotic
  behavior~\cite{Berge84}.  Properties of a burst such
  as its number of spikes and its set of interspike
  intervals will evolve over time, with the transient
  time depending on the choice of initial condition
  (how far the initial state is from the attractor) and
  on neuronal parameters.  Our results such as
  Fig.~\ref{fig:initCondProp} for an homogeneous
  HH~neuronal model suggest that transient times can in
  fact be rapid, just four or fewer successive neurons.
  Fig.~2 of Hahnloser et al~\cite{Hahnloser02} does not
  show evidence of transient behavior since later
  bursts do not seem to be statistically different from
  earlier bursts during the same motif. However, only a
  small number (about~20) of~\hvcra~neurons have been
  sampled to date in singing birds, and neurons that
  burst at the beginning of a motif have not been found
  so further study involving more neurons would be
  worthwhile. It may also be the case that the songbird
  brain has evolved in such a way as to reduce or
  eliminate the duration of transients, for example by
  starting a chain with a state close to the asymptotic
  burst, or by using inhibitory neurons to accelerate
  convergence to synchronized firing as bursts
  propagate from projection neuron to projection
  neuron~\cite{Kopell04}.
  
\item A one dimensional chain is not robust since
  propagation terminates if any neuron in the chain
  dies.  A more realistic synfire chain must have at
  least two neurons in a pool to be robust, and the
  neurons in a pool must synchronize their spikes to
  some extent to achieve reliable transmission from
  pool to pool~\cite{Abeles91}. A synfire chain
  hypothesis thus predicts that there must be more than
  one \hvcra~neuron bursting at any give time (these
  are the neurons that belong to the same pool) and
  further that neurons that burst at about the same
  time should have nearly synchronous bursts. Thus
  Fig~2 of Ref.~\cite{Hahnloser02} may be incomplete
  and a high-resolution optical study of~HVC (or
  possibly a many-electrode study) during the singing
  of a motif or during the audition of a bird's own
  song~\cite{Mooney00} may reveal multiple
  \hvcra~neurons firing bursts in synchrony.
  
  Abeles has pointed out that there is an inverse
  relation between the number of neurons needed in a
  synfire pool for reliable propagation and the average
  strength of the synapses between
  pools~\cite{Abeles91}. Using paired intracellular
  recordings, Mooney and Prather~\cite{Mooney05} have
  shown that synapses between \hvcra~neurons are
  stronger than mammalian cortical neurons by about an
  order of magnitude (the average~EPSP magnitude is
  about 2~mV compared to about 0.2~mV in cortex) so
  chains in~HVC would be expected to have about an
  order of magnitude fewer neurons in a pool. Given
  Abeles's estimate of about 50-100~neurons in a
  cortical pool, perhaps ten or fewer neurons might be
  needed in an HVC~synfire pool.  Assuming that the
  HVC~bursts of average duration 6~ms are
  non-overlapping, a zebra finch motif of duration
  0.6~s would require about $0.6/0.006 \approx 100$
  successive pools to span the motif, so a synfire
  chain in~HVC might involve about $100 \times 10
  \approx 1,000$ neurons. This is a much smaller number
  than the 40,000 \hvcra~neurons estimated to exist
  in~HVC. This suggests that a synfire chain might be
  too simplistic an architecture to justify the large
  number of \hvcra~neurons.


\item 
  It is worth noting that a synfire chain architecture
  can explain how an overall precise timing can be
  maintained in~HVC even though there is a steady and
  substantial turnover of neurons in~HVC throughout the
  life of the songbird~\cite{NottebohmJN02,Wang02}.
  Since perfect synchronization of a pool is not needed
  to guarantee transmission of information through the
  next pool, a fraction of neurons in a given pool can
  alter their properties or fail without degrading the
  transmission of information or its timing.
  
\end{enumerate}

In conclusion, our calculations support the hypothesis
that the sparse bursts observed experimentally in
\hvcra~neurons during singing can be understood as the
propagation of bursts through an excitatory feedforward
synfire chain. However, further experiments and
computational studies are needed to confirm this
hypothesis and to rule out the competing hypotheses
discussed in
Section~\ref{sec:other-hypothese-and-related-theory}.
Especially interesting in the near term would be to
understand some of the quantitative details of the
experiments, for example the production of a brief 6~ms
burst of four spikes corresponding to a frequency of
about 600~Hz that is consistent with the known
properties of HVC~neurons and of the HVC
microcircuitry~\cite{Mooney05}. Although our results
show that a brief burst can propagate in a stable
manner, we and others~\cite{Jin05} have not been able
to construct neuronal models, or simple networks of
such models, that burst as rapidly and as briefly as
actual \hvcra~neurons.

\vspace{.2in}

Note: Toward the completion of this paper, the authors
received a preprint by Jin et al whose results overlap
with the present paper (the preprint is related to a
recent conference presentation~\cite{Jin05}). These
authors also investigated whether brief high-frequency
bursts analogous to those observed in \hvcra~neurons
could propagate through a purely excitatory network,
and they too found that this was possible for a range
of parameters, thereby supporting the hypothesis that
the bursting of \hvcra~neurons could be intrinsic
to~HVC and causal.

The main difference of the present paper with the
preprint is that we explored whether bursts could
propagate under the simplest circumstances of a 1d,
homogeneous, noiseless, purely-feedforward chain using
simple neuronal models.  Jin et al studied a more
complicated generalized (not purely feedforward),
heterogeneous, noisy, purely excitatory synfire chain
with many neurons per pool. Jin et al also used a more
complicated two-compartment conductance-based neuronal
model that included a low-threshold potassium channel
that provided a strong spike-frequency adaptation
similar to that observed experimentally. (This helped
to create brief high-frequency bursts.) They also
included calcium dynamics in the dendrite compartment
that could lead to the firing of calcium spikes. Jin et
al were able to show that adding the low-threshold
potassium channel and calcium dynamics made the
propagation of bursts through their generalized synfire
chain more robust (one did not have to tune parameters
so carefully to obtain a stable propagating
burst). They also studied how jitter in the burst
evolved along the length of the synfire chain, which we
have also done in a forthcoming paper with a somewhat
different model~\cite{Li06II}.

Given that our calculations and those of Jin et al have
left out the role of inhibitory neurons and were based
on a limited knowledge of~HVC neuronal properties and
of the HVC microcircuitry (e.g., there is no evidence
yet for a synfire chain in HVC, nor have the
conductances of HVC~neurons been fully characterized),
both calculations should be considered as modest but
complementary and useful steps toward understanding
possible mechanisms for the dynamics of~HVC during
singing.

\appendix

\section{Hodgkin-Huxley Equations and Parameters}

\label{appendix:equations-parameters}

The single-compartment HH~model of
Eq.~(\ref{eq:HH-model}) had the following five
representative currents~$I_i(t)$:
\begin{eqnarray}
  I_{\rm leak} &=& \bar{g}_L (v-E_L), \nonumber \\
  I_{\rm Na} &=& \bar{g}_{\rm Na} m^3 h (v-E_{\rm Na}), \nonumber\\
  I_{\rm K} &=& \bar{g}_{\rm k} n^4 (v- E_{\rm K}),   \label{eq:hh-currents}
 \\
  I_{\rm Ca} &=& \bar{g}_{\rm Ca} m_1^2 h_1 (v-E_{\rm Ca}),\nonumber \\
  I_{\rm KCa} &=& \bar{g}_{\rm KCa} n_1^3 (v-E_{\rm KCa}),\nonumber
\end{eqnarray}
with maximum conductances per unit area y (units of
$\rm mS/mm^2$) given
\begin{equation}
\label{eq:hh-max-gbar-values}
  \bar{g}_L = 0.003, \quad
  \bar{g}_{\rm Na} = 1.0 , \quad
  \bar{g}_{\rm K} = 0.22 , \qquad
  \bar{g}_{\rm Ca} = 0.031 , \qquad
  \bar{g}_{\rm KCa} = 0.007 , \qquad
\end{equation}
and corresponding reversal potentials (units of~mV):
\begin{equation}
\label{eq:hh-rev-potentials}
  E_L = -70, \quad
  E_{\rm Na} = 50 , \quad
  E_{\rm K} = -80 , \qquad
  E_{\rm Ca} = 120 , \qquad
  E_{\rm KCa} = -80 . \qquad
\end{equation}

\begin{table}[tbh]
  \label{table:hh-parameters}
  \begin{tabular}{c|c|c|c|c|c|c|}
     \multicolumn{1}{c}{} 
   & \multicolumn{1}{c}{\mbox{\ \ \ $m_{\ }$\ \ \ } }
   & \multicolumn{1}{c}{\mbox{\ \ \ $n_{\ }$\ \ \  } }
   & \multicolumn{1}{c}{\mbox{\ \ \  $h_{\ }$\ \ \  } }
   & \multicolumn{1}{c}{\mbox{\ \ \  $m_1$\ \ \ }}
   & \multicolumn{1}{c}{\mbox{\ \ \ $h_1$\ \ \  }}
   & \multicolumn{1}{c}{\mbox{\ \ \ $n_1$\ \ \  }}\\
    $v_0$ mV \, & -25.5 & -12.3  &  -60  & -50 & -80 &  \\
    $v_1$ mV \, & 5.29 & 11.8 & -15 & 5 & -5 &  \\
    $v_2$ mV \, & -120 & -28.3 & -62.9 & -50 & -80 & -35 \\
    $v_3$ mV \, & 25 & 19.2 & 10 & 10 & 10 & 10 \\
    $t_1$ ms \, & -1.26 & -6.4 & 1.1 & 20 & 60 & 30\\
    $t_2$ ms \, & 1.32 & 7.2 & 0  & 10 & 10 & 10
  \end{tabular}
  \caption{Parameter values for the Hodgkin-Huxley model
         Eq.~(\ref{eq:HH-model}) with currents
  Eq.~(\ref{eq:hh-currents}) and gating equations
  Eqs.~(\ref{eq:x-sigmoid-fn})-(\ref{eq:ca-tau-form}).
  }
\end{table}

The currents Eqs.~(\ref{eq:hh-currents}) depend on the
membrane voltage~$v(t)$ and on gating variables $m$,
$h$, $n$, $m_1$, $h_1$, and~$n_1$ that determine the
probability as a function of time for certain channel
subunits to be open. All the gating variables~$x(v)$
except the KCa activation variable~$n_1$ obey the
evolution equation
\begin{equation}
  \label{eq:gating-variable-eq}
  \tau(v) \frac{dx}{dt} = x_{\infty}(v) - x  ,
\end{equation}
with the asymptotic value~$x_\infty(v)$ having the
sigmoidal form
\begin{equation}
  x_{\infty}(v) = \left( 
    1 + e^{\displaystyle -(v-v_0)/v_1} \right)^{-1} .
  \label{eq:x-sigmoid-fn}
\end{equation}
The time constant~$\tau(v)$ has the form
\begin{equation}
  \tau(v) = t_2 + 
  t_1 \left(
    1 + e^{\displaystyle
    -\left( (v-v_2)/v_3 \right) }
   \right)^{-1} .
   \label{eq:hh-tau-fn}
\end{equation}
for the standard HH channels~$m$, $h$, and~$n$, and the form
\begin{equation}
  \label{eq:ca-tau-form}
  \tau(v) = t_2 + 
    t_1 \exp\left[  
      -\left( \frac{v-v_2}{v_3} \right)^2 
    \right] ,
\end{equation}
for the Ca channels~$m_1$ and~$h_1$ and KCa channel $n_1$.
Table~\ref{table:hh-parameters} gives the parameter
values  that we used in these various expressions.

The KCa~activation variable~$n_1$ obeys the evolution
equation
\begin{equation}
  \tau_{\rm KCa} \frac{dn_1}{dt} = f\left([{\rm
  Ca}]_i\right)-n_1 ,
\end{equation}
where the expression~$[{\rm Ca}]_i$ denotes the average
intracellular calcium concentration and the
function~$f$ is given by
\begin{equation}
  f(\cai) = \frac{\cai}{(\cai+0.001)(1+\exp(-(v+40)/20))} .
\end{equation}
The intracellular calcium concentration accumulates
when the low threshold calcium channel is open and
calcium ions flow from extracellular solution into
intracellular solution.  The phenomenological model
for~$\cai$ was taken from Ref.~\cite{Turrigiano1995}
\begin{equation}
  \cai = ([{\rm Ca}]_{i0} - \cai - F*(I_{ca}))/\tau_{Ca} ,
\end{equation}
where the equilibrium Calcium concentration~$[{\rm
  Ca}]_{i0}$ had the value~.00005~mM and the
current-density factor~$F$ had the value 3~${\rm
  mM-cm}^2/{\rm mA}$, and the time constant~$\tau_{Ca}$
had the value~200 ms.  The KCa channel was chosen to
imitate a low-threshold transient calcium current
observed in HVC neurons by Kubota~\cite{Kubota98}.

\begin{acknowledgments}
  The authors thank Richard Mooney, Jonathan Prather,
  Melissa Coleman, Michale Fee, Stephen Shea, George
  Augustine, and Paul Tiesinga for helpful discussions,
  and Craig Henriquez and Duke University's Center for
  Neural Engineering for providing support for one of
  the authors (M.~Li).
\end{acknowledgments}

\end{document}